\documentclass[11pt]{article}  %\Box not defined
\def\ds{\displaystyle}

\begin{document}
\pagestyle{headings}
\renewcommand{\thefootnote}{\alph{footnote}}

\title{Measurement and self-adjoint operators}
\author{Glenn Eric Johnson\\Oak Hill, VA.\footnote{Author can be reached at: glenn.e.johnson@gmail.com.}}
\maketitle

{\bf Abstract:} The approximations of classical mechanics resulting from quantum mechanics are richer than a correspondence of classical dynamical variables with self-adjoint Hilbert space operators. Assertion that classical dynamic variables correspond to self-adjoint Hilbert space operators is disputable and sets unnatural limits on quantum mechanics. Well known examples of classical dynamical variables not associated with self-adjoint Hilbert space operators are discussed as a motivation for the realizations of quantum field theory that lack Hermitian field operators but exhibit interaction. 


% = = = = = = = = = = = = = = = = = = = = = = = = = = = = = = = = = = = = = = = =

%introduction
% = = = = = = = = = = = = = = = = = = = = = = = = = = = = = = = = = = = = = = = =
\section{Introduction}

Quantum mechanics provides a versatile description of nature. This description is more general than the canonical quantizations successfully used to describe nonrelativistic quantum dynamics. Canonical quantizations largely preserve the intuition derived from classical mechanics that dynamics results from consideration of geometric objects following trajectories governed by differential equations. This conjecture that quantum mechanics results from association of self-adjoint operators in a Hilbert space with the dynamical variables describing geometric objects evolving in a configuration space has nonrelativistic quantum mechanics and Feynman rules series as exemplars, although the association for the Feynman rules is formal. However, there are difficulties with this approach, particularly when relativity is considered. Canonical quantization extrapolates classical descriptions to quantum mechanical settings that generally lack a characterization as identifiable objects traveling on trajectories. Consideration of Young's double slit and resolution of Gibbs' paradox clarify that quantum mechanics does not describe distinguishable objects traveling trajectories. Ehrenfest's theorem provides justification for the correspondence of classical and quantum dynamical variables in the case of nonrelativistic physics and special circumstances but any correspondence once particle production becomes likely is curious. This curious extrapolation and the difficulty of demonstrating realizations of interest for quantum field theory (QFT) motivates examination of alternatives to canonical quantization, in particular, alternatives to the conflation of dynamic variable with self-adjoint operator. As the more general case, quantum mechanics should stand on its own [\ref{longo}].

This note discusses the description of quantum mechanics with an emphasis on states as elements of Hilbert spaces. The emphasis on the elements of the Hilbert space rather than conjecture for operators frees quantum mechanics from inconsistent assertions. This note argues that canonical quantization is inappropriate for relativistic quantum mechanics and discusses the Everett-Wheeler-Graham (EWG) relative state interpretation [\ref{ewg}] that does not rely on classical concepts to complete quantum mechanics. The first discussion emphasizes that many quantities of interest are not self-adjoint operators in Hilbert spaces with an infinite number of dimensions. The second discussion emphasizes the demonstration that quantum mechanics is a complete and consistent model of nature. In particular, the EWG interpretation rids quantum mechanics of reliance on a classical domain and the ad hoc process of wave packet collapse to the eigenvectors of the operations associated with measured quantities. Descriptions of quantum mechanics that assume self-adjoint Hilbert space operators correspond to classical dynamical variables are not adequate in rigged Hilbert spaces. The association of observable quantities with self-adjoint operators is a relatively intuitive formulation for quantum mechanics. It is more general to consider that the physical states are elements of appropriate Hilbert spaces and that observables are characterizations of those states. Consideration of only the states, necessarily realized as elements in the Hilbert spaces, together with an EWG interpretation results in a more general description of quantum mechanics.

Questions of interpretation are substantial and affect the mathematical formulation of quantum mechanics when they distinguish observable, meaning susceptible to measurement, from an {\em observable}, meaning a self-adjoint Hilbert space operator. Failure of self-adjointness for an operation corresponding to a classical dynamical variable does not exclude the quantity from the characterizations of states. The question of interpretation arises since canonical quantization relies on precise associations of Hilbert space operators and classical dynamical variables. Quantum mechanics accommodates more general descriptions and the precise associations used in canonical quantizations are disputable.

% = = = = = = = = = = = = = = = = = = = = = = = = = = = = = = = = = = = = = = = =

%established description of quantum mechanics

% = = = = = = = = = = = = = = = = = = = = = = = = = = = = = = = = = = = = = = = =
\section{Quantum mechanics}

\subsection*{Canonical quantization-based and general descriptions}

Inquiry into the appropriate realizations for quantum mechanics was initiated notably by John von Neumann.  This inquiry into the appropriate formulation for the quantum mechanics described in the pioneering work of Erwin Schr\"{o}dinger, Werner Heisenberg, P.A.M.~Dirac, Pascual Jordan, Max Born, Wolfgang Pauli and Eugene Wigner was extended by Arthur Wightman, Rudolf Haag, Huzihiro Araki, Res Jost and Hans-J\"{u}rgen Borchers whose works and [\ref{yngvason-lqp}] should be consulted for background and many other significant contributors. The description uses the concept of Hilbert space notably due to David Hilbert, Erhard Schmidt and Frigyes Riesz. The discussion here is intended to emphasize a perspective on operators and measurement that is decisive to realizations of QFT with interaction in physical spacetimes. Examples [\ref{gej05},\ref{mp01},\ref{feymns}] demonstrate that difficulties in QFT can be attributed to inclusion of unnecessary assertions within the foundations. Dynamical description beyond canonical quantization may be necessary to a consistent development of relativistic quantum mechanics that exhibits appropriate classical limits.
 
Established descriptions of quantum mechanics, for example [\ref{dirac},\ref{bjdrell},\ref{weinberg}], include:
\newcounter{mycount}
\begin{list}{P\arabic{mycount}.}%
{\usecounter{mycount}}
\item\label{it-state} {\em States.} States $\rho$ are positive semi-definite density matrices with a normalized trace, also designated statistical operators [\ref{vonN},\ref{gleason}]. Trace($\rho)=1$. A vector state is described by an element $|\Psi\rangle$ of the Hilbert space and then $\rho$ is the projection onto $|\Psi\rangle$.
\item\label{it-time} {\em Dynamical evolution.} Time evolution is determined by unitary transformation $U(t)$.\[\rho(t) = U(t)\rho(0)U(t)^{-1}.\]For the vector states, time evolution follows the Schr\"{o}dinger equation involving the self-adjoint generator of time translations, the Hamiltonian, $H$,\[-i\hbar \frac{\partial |\Psi\rangle}{\partial t}=H|\Psi\rangle.\]$U(t)=\exp(iHt/\hbar)$.
\item\label{it-obs} {\em Measurement.} Dynamical variables are self-adjoint Hilbert space operators $A$. The vector states are linear combinations of the eigenvectors of $A$, $|\Psi \rangle = \sum a_\lambda |\lambda\rangle$ and $A| \lambda\rangle = \lambda |\lambda \rangle$. The mean (expectation) value of an observable for the state $\rho$ is Trace$(A \rho)$. For a vector state $|\Psi\rangle$, Trace$(A\rho)=\langle \Psi|A\Psi \rangle=\sum \lambda |a_\lambda|^2$. The likelihood that the result associated with the vector state $|\lambda\rangle$ is observed is Trace($P_\lambda \rho$) with $P_\lambda$ the projection onto $|\lambda \rangle$. When the value $\lambda$ is observed in a measurement of the dynamical variable $A$ and the solutions to $A| \lambda\rangle = \lambda |\lambda \rangle$ are a one dimensional subspace, then the state is subsequently described by the time evolution of the corresponding eigenvector $|\lambda\rangle$.
\end{list}For the discussion below, P\ref{it-state}-P\ref{it-obs} are taken as the canonical quantization-based description of quantum mechanics. The change in emphasis below is to note that the self-adjointness condition from P\ref{it-obs} limits the definition of {\em observable} but does not constrain the dynamical variables used to characterize the states. All normal real linear operators are self-adjoint in finite dimensional Hilbert spaces when all linear operators are bounded and all domains are the entire Hilbert space. But in more general Hilbert spaces, P\ref{it-obs} excludes essential dynamical variables such as position and field strength as observables. P\ref{it-obs} is restated below for compatibility with more general Hilbert spaces and to eliminate unnecessary conjecture.

P\ref{it-obs} includes three distinct assertions for dynamical variables:\begin{enumerate}\item dynamical variables describe the states \item dynamical variables correspond to self-adjoint operators \item the eigenvectors of the corresponding self-adjoint operators are states.\end{enumerate}Here {\em states} refers to elements of the Hilbert space. Use of dynamical variables to characterize the states is self-evident. Location, energy-momentum, charge and spin are observable properties used in descriptions of states. However, the assertions 2 and 3 fail in cases of interest [\ref{vonN}]. In contradiction to P\ref{it-obs}, dynamical variables that are not associated with self-adjoint operators or have eigenvectors that are not states have appeared in the practice of quantum mechanics since its inception. For example, in the ${\cal L}_2$ Hilbert space of ordinary (nonrelativistic) quantum mechanics, functions of $x$ and functions of $p$ are self-adjoint operators but their products are not necessarily self-adjoint. That symmetric products of self-adjoint operators are not necessarily self-adjoint demonstrates one inadequacy of the correspondence of classical and quantum mechanical dynamical variables [\ref{vonN}]. Position $x$ is an example of a dynamical variable that due to Lorentz invariance lacks an associated self-adjoint operator [\ref{wigner}] and the eigenvectors of $x$, $|x\rangle$, are not states. In the Hilbert space of square-integrable functions ${\cal L}_2$,\[P_{a,b}:=\int_a^b dx \; |x\rangle \langle x|\]formed from eigenstates of location $|x\rangle$ are Hilbert space projections with the definition\[\langle v | P_{a,b} u\rangle := \int_a^b dx \; \overline{v(x)} u(x),\]but $P_{a,a}=0$ and not the projection onto the location eigenstate $|a\rangle$. The evident meaning of the notation $|x\rangle \langle x|$ is not as a Hilbert space projection operator when $|x\rangle$ is not an element of the Hilbert space. A more general formulation of the the bra and ket notation of [\ref{dirac}] uses rigged Hilberts spaces with Gelfand triples, spaces of generalized functions that are dual to spaces of test functions and contain rigged (equiped) Hilbert spaces as completions of the spaces of test functions. The spectral (kernel) theorem [\ref{gel4}] provides that a self-adjoint operator is a real linear combination of projection operators but the eigenvectors of the self-adjoint operator may be generalized eigenvectors rather than elements of the Hilbert space. The limited class of dynamical variables that correspond to self-adjoint operators with eigenvectors that are elements of the Hilbert space does not include essential dynamical variables used to describe states in quantum mechanics.

In P\ref{it-obs}, collapse of states upon observation associates states with classically idealized eigenvectors. These eigenvectors satisfy $A|\lambda\rangle=\lambda\,|\lambda\rangle$ for the operation $A$ associated with the dynamical variable, for example, for position $x$ the eigenstate is a Dirac delta (generalized) function. Below, no elaboration of quantum mechanics with a classical domain is considered nor is quantum mechanics considered a statistical theory for underlying objects that are described by classical idealizations. The changes to a canonical quantization-based description of quantum mechanics are implemented by considering only those Hilbert space operators realized in particular Hilbert spaces and the Everett-Wheeler-Graham (EWG) relative state interpretation of quantum mechanics. Self-adjoint projection operators onto the elements of the Hilbert space and generators for the Poincar\'{e} symmetries are necessarily realized. This discussion focuses on explicit Hilbert space realizations and [\ref{yngvason-lqp}] should be consulted for an algebraic development.

Characterizations of states include interpretations by an observer who applies the classical idealizations. In particular, a state that is arbitrarily predominantly within a small region may be interpreted as being located at a point. This interpretation of the state is justified with neither the state being an eigenvector of a location operator $x$ nor $x$ being realized as a self-adjoint operator in the Hilbert space. States dominantly supported in small regions but not even of bounded support may be interpreted as being located at a point. This characterization is physically justified since observations of the state beyond the small region can be exceedingly rare. The examples of a relativistic position operator [\ref{wigner}], and the contradictions to QFT defined at a point [\ref{wizimirski},\ref{wight-pt}] clarify difficulties with classical idealization. To satisfy Poincar\'{e} invariance, VEV can not be considered as functions and states are labeled by test functions rather than points in configuration space. Observers typically label results with points in configuration space.

Dynamical variables are observable in the sense that they correspond closely with particular states although they may not be {\em observables} as described by P\ref{it-obs}. The promotion of classical dynamical variables to Hilbert space operators provides a classically intuitive translation of classical mechanics to quantum mechanics but such an association is not necessary and is problematic, particularly with the inclusion of relativistic invariance [\ref{yngvason-lqp},\ref{wigner},\ref{segal}]. A description of dynamical variables consistent with a quantum mechanics that includes representations of the Poincar\'{e} group and eigenstates of position is:\begin{enumerate} \item[P\ref{it-obs}'.] {\em Measurement.} States are described using dynamical variables. The likelihood that the result associated with an element $|\Psi\rangle$ is observed for a state described by the density matrix $\rho$ is Trace($P_\Psi \rho$) with $P_\Psi$ the orthogonal projection onto $|\Psi\rangle$. For a vector state $|\Psi'\rangle$, Trace$(P_\Psi \rho)=\langle \Psi'|P_\Psi \Psi' \rangle=|\langle \Psi'|\Psi\rangle|^2$.\end{enumerate}Dynamical variables derive from the arguments in a representation of a generalized function as a summation (Section 4.1, Chapter II of [\ref{gel2}]). States with a finite number of discrete values are included as degenerate cases, and fields would be considered as functional integrals. 

P\ref{it-obs}' generalizes P\ref{it-obs} to include rigged Hilbert spaces and avoids the constraints of measurement process conjecture. P\ref{it-obs}' eliminates the unnecessary assertion of state collapse subsequent to measurement and eliminates the definition of an observable as a self-adjoint Hilbert space operator. The P\ref{it-obs} description of state collapse is limited to perfect measurement in a sense discussed below. P\ref{it-obs}' takes Born's rule as axiomatic and limited to states. Orthogonal projection operators onto states are inherent to Hilbert space [\ref{rudin}]. The anticipated collapse upon measurement is virtual in the EWG interpretation of quantum mechanics, and the collapse is not necessarily associated with eigenvectors of the operation associated with the observed quantity. P\ref{it-obs}' is suggested as a working condition to mature into an axiom.

Projection operators provide a natural setting for description of quantum mechanics [\ref{vonN},\ref{birk}]. In rigged Hilbert spaces (Gelfand triples), the kernel (spectral) theorem [\ref{gel4}] provides expansions of self-adjoint operators as linear combinations with real coefficients of projection operators $P_\theta$ from a resolution of unity ($\sum_\theta\; P_\theta=1$). The $P_\theta$ include both ``projections'' onto subspaces of states and generalized states such as eigenvectors for position and momentum that are duals of test functions but not elements of the Hilbert space. Assertion that states are linear combinations of states with classical characteristics, for example, linear combinations of eigenvectors of self-adjoint position or field strength operators, is not general. In separable Hilbert spaces, Hilbert space projection operators have a one-to-one correspondence with subspaces of states [\ref{rudin}] and the self-adjoint operators are weighted summations of Hilbert space projection operators [\ref{gel4}] but the eigenstates of self-adjoint operators are not necessarily associated with Hilbert space projections. Thus, collapse of the states to classically idealized states such as eigenvectors for position, momentum or field strength is not supported in general Hilbert spaces.

The emphasis of P\ref{it-obs}' is on the elements of the Hilbert space and observer's characterizations of those elements, characterizations that include interpretation using idealized classical dynamical variables. Canonical quantization adds conjecture concerning the realization of operators and eigenvectors. In P\ref{it-obs}', the dynamical variables are required only to describe the states. The limitations of assertions 2 and 3 for {\em observables} are widely known despite the use in statements of the principles of quantum mechanics. 2 and 3 describe a classically idealized, statistical interpretation of quantum mechanics more appropriate in finite dimensional Hilbert spaces.

% = = = = = = = = = = = = = = = = = = = = = = = = = = = = = = = = = = = = = = = =

%examples of defects to description in Hilbert spaces of qm

% = = = = = = = = = = = = = = = = = = = = = = = = = = = = = = = = = = = = = = = =
\subsection*{Defects of self-adjoint Hilbert space operators as dynamical variables}

Dynamical variables that are not associated with self-adjoint operators are safely within conventional descriptions of quantum mechanics. Two examples of this are $x^3 p$ in ${\cal L}_2$ [\ref{bogo}] and $x$ in the one particle subspace of Fock space [\ref{wigner}]. Established axioms for QFT make assumptions for operators, for example, the G\aa rding-Wightman axioms assert that fields are Hermitian Hilbert space operators with dense domains. The examples of $x$ and $x^3p$ illustrate that there are alternatives to Hermiticity within quantum mechanics, and the examples are counterexamples to assertion that dynamical variables are necessarily self-adjoint Hilbert space operators. 

The $x^3p$ example is for ordinary quantum mechanics with a single degree of freedom. $x^3 p$ corresponds to the formally self-adjoint\[-i\hbar x^{3/2}\frac{d}{dx}x^{3/2}=-i\hbar(\frac{x^3}{2\;}\frac{d}{dx}+\frac{d}{dx} \frac{x^3}{2\;})\]and has square summable, normalized eigenfunctions\begin{equation}\label{ce-sa} s_\lambda(x):= \sqrt{2\lambda}\;\; \frac{\exp{(-\lambda/(2x^2))}}{x^{3/2}}\end{equation}with imaginary eigenvalues $-i\hbar \lambda$ [\ref{bogo}]. The $s_\lambda(x)\in {\cal L}_2$ provided here is for $x\geq 0$ and $s_\lambda(x)=0$ for negative $x$. The operator associated with $x^3 p$ is evidently not self-adjoint for ${\cal L}_2$ although $x^3p$ is well defined in classical dynamics. Nevertheless, for minimum uncertainty packet states $s_t(x)$ with small spatial variances, the trajectory of $x^3p$ given by Newtonian mechanics approximates $\langle s_t| X^{3/2}PX^{3/2} s_t\rangle$ from quantum dynamics [\ref{schrodinger}]. This establishes that there are particular states with real $\langle s_t| X^{n/2}PX^{n/2} s_t\rangle$ that agree with classical limits $x^np$ even though $X^{n/2}PX^{n/2}$ is not self-adjoint. Assertion that classical dynamical variables correspond to self-adjoint operators is unjustified without additional conditions. And, even when a classical quantity lacks an associated self-adjoint operator, classical limits of quantum mechanics may include good approximations of the classical dynamics.

In ${\cal L}_2$, the Riesz-Fischer theorem [\ref{rudin}] provides that the eigenfunctions of $x$ and $p$ are complete. $x$ or $p$ are associated with resolutions of unity provided by Lebesgue measure and the Fourier transform. Products of $x$ and $p$ can lack corresponding self-adjoint operators as demonstrated by the example of $x^3p$. Quantities such as $x^3p$ that are dynamical variables in classical mechanics are excluded as self-adjoint operators in ${\cal L}_2$.

% \section{A case study, the harmonic oscillator}

The linear harmonic oscillator provides an example with the quantum dynamics providing a good approximation for the classical dynamics for a quantity $x^3p$ that lacks a corresponding self-adjoint operator. The linear harmonic oscillator was used by Schr\"{o}dinger to study the correspondence of quantum with classical mechanics. In the ordinary quantum mechanical description, the states of a particle are ${\cal L}_2$ equivalence classes labeled by functions $f(x)$. In the ordinary (nonrelativistic) quantum mechanical description, time is an independent parameter and the position and momentum operators satisfy the Heisenberg-Born-Jordan relation, $[P,X]=-i \hbar$, realized in ${\cal L}_2$ by $X=x$ and $P=-i\hbar \frac{d\;}{d x}$. Time evolution satisfies the Schr\"{o}dinger equation for the linear harmonic oscillator Hamiltonian\begin{equation} \label{hohamil} H=-\frac{\hbar^2}{2m} \frac{d^2\;}{d x^2} +\frac{k}{2}\,x^2\end{equation}described by a tension constant $k$ and the particle mass $m$.

Minimum packet states are the states most nearly described as classical states in the sense that the geometric mean of the variances in position and momentum are minimal. There are solutions to the Schr\"{o}dinger equation for the harmonic oscillator that are minimum packet states.\begin{equation} \label{minpak} s_t(x)=\frac{1}{\sqrt{2\pi \sigma^2}}\; \exp\left(-\frac{(x-A \cos w t)^2}{4\sigma^2}-i \frac{\beta x \sin wt}{\hbar} -i\phi(t) \right)\end{equation}with\[ \renewcommand{\arraystretch}{1.25} \begin{array}{l} \sigma^2 = \frac{\ds \hbar}{\ds 2\sqrt{mk}}\\ 
 \beta = \sqrt{mk}\, A\\ 
 w = \sqrt{k/m}\\
 \phi(t)= \frac{\ds w}{\ds 2} t - \frac{\ds k A^2 }{\ds 4 \hbar w} \sin 2wt. \end{array} \]For these states and with $\langle T \rangle_t := \langle s_t | T s_t\rangle$,\begin{equation}\label{schro-soln} \renewcommand{\arraystretch}{1.25} \begin{array}{l} \langle X\rangle_t= A \cos wt\\
\langle P\rangle_t= -\sqrt{mk}\, A \sin wt\\
\langle (X-\langle X\rangle_t )^2 \rangle_t= \sigma^2\\
\langle (P-\langle P\rangle_t)^2 \rangle_t= \hbar^2/(4\sigma^2)\\
\sqrt{\langle (X-\langle X\rangle_t)^2 \rangle_t \langle (P-\langle P\rangle_t)^2 \rangle_t}=\hbar/2, \end{array}\end{equation}the minimum consistent with the Heisenberg uncertainty. The quantity of interest is\[\langle X^{3/2}PX^{3/2} \rangle_t = -\sqrt{mk}\, A\sin wt(A^3 \cos^3 wt +3A\sigma^2 \cos wt)\]for the states $s_t(x)$. A classical limit should apply when the variance $\sigma^2 \ll A^2$ and in particular as $m \rightarrow \infty$\[\sigma^2 =\frac{\ds \hbar}{\ds 2\sqrt{mk}} \ll A^2= \frac{\ds 2E}{\ds k}\]with the total energy $E$ a constant of the motion.

The classical description is that a point mass evolves along a trajectory $x(t)$. The trajectory is described by Newton's equation of motion.\begin{equation} m \ddot{x} = -k x\end{equation}with the solution\begin{equation}\renewcommand{\arraystretch}{1.25} \begin{array}{rl} x&=x_o \cos(w t)+(p_o/\sqrt{km})\, \sin(w t)\\
 &=A \cos(w t +\theta)\\
w &=\sqrt{k/m}\\
x_o&=x(0)=A\cos(\theta)\\
p_o&=m \dot{x}(0)=-A\sqrt{km}\sin(\theta).\end{array}\end{equation}With $\theta=0$, $x_o=A$ and $p_o=0$, the classical limit agrees with the quantum dynamics given in (\ref{minpak}) and (\ref{schro-soln}) as anticipated by Ehrenfest's theorem for ordinary quantum mechanics. For these minimum packet states and the harmonic oscillator,\[\renewcommand{\arraystretch}{1.25} \begin{array}{rl} x&= \langle X \rangle_t\\ p&= m \dot{x} =\langle P \rangle_t\\ x^3 p &= A^3 \cos^3 wt \,(-\sqrt{mk}\, A\sin wt)\\
\langle X^{3/2}PX^{3/2} \rangle_t &= -\sqrt{mk}\, A\sin wt(A^3 \cos^3 wt +3A\sigma^2 \cos wt).\end{array}\]When the variance $\sigma^2 \ll A^2$ and in particular as $m \rightarrow \infty$,\[x^3 p \approx \langle X^{3/2}PX^{3/2} \rangle_t.\]For the minimum packet states in a classical limit, the detailed dynamical results for the harmonic oscillator (\ref{schro-soln}) approximate the trajectory $x(t)$. For these minimum packet states and large masses, there is a close correspondence of quantum dynamics with the classical dynamics.

For the states (\ref{ce-sa}) that demonstrate that $x^3p$ does not correspond to a self-adjoint operator,\[\renewcommand{\arraystretch}{1.25} \begin{array}{l} \langle s_\lambda |X s_\lambda \rangle = \sqrt{\pi \lambda}\\
 \langle s_\lambda |P s_\lambda \rangle = 0\\ \langle s_\lambda | X^{3/2}PX^{3/2} s_\lambda \rangle = -i\hbar \lambda. \end{array}\]$\langle s_\lambda |H s_\lambda \rangle$ diverges for (\ref{hohamil}) but this is not the decisive consideration as the expected value of the Hamiltonian is finite for (\ref{hohamil}) and eigenvectors of $x^np$ with imaginary eigenvalues when $n\geq 4$. For the states $s_\lambda$, the quantum dynamics lacks an evident correspondence with the classical trajectory.

The operator corresponding to the classical quantity $x^3p$ is not a self-adjoint operator even though $x^3p$ is well approximated by quantum dynamics for the minimum packet states. Of course, when packet spread is appreciable, expectation values do not necessarily follow the classical dynamical descriptions. For the example of the eigenvectors $s_n(x)$ of the Hamiltonian for the harmonic oscillator (\ref{hohamil}), the spatial spreads are appreciable with respect to the amplitude $A$. For these eigenvectors,\[\sigma^2 :=\langle s_n| X^2 s_n \rangle = (n+{\textstyle \frac{1}{2}})\hbar /(mw)= {\textstyle \frac{1}{2}}A^2\]from $E=(n+\frac{1}{2})\hbar w = \frac{1}{2}m w^2 A^2$ and $\langle s_n|X s_n\rangle=\langle s_n|P s_n\rangle =0$. Validity of the classical limit relies both on $m\rightarrow \infty$ and the selection of appropriate states.

The next example is the lack of a self-adjoint location operator for the relativistically invariant scalar product in the one particle subspace of Fock space for a massive, Lorentz scalar free field. With Lorentz invariance, position is a property of states that lacks a corresponding self-adjoint operator [\ref{wigner}] and projections onto bounded regions are inconsistent with causality [\ref{yngvason-lqp}]. Multiplication by a spatial component of $x$ is generally complex.\[\renewcommand{\arraystretch}{1.25}\begin{array}{rl} \langle f| {\bf x} g\rangle &= {\ds \int} dxdy\; \Delta (x-y) \overline{f(y)}\, {\bf x}\, g(x)\\
 &= i\hbar {\ds \int} dp \;{\ds \frac{\delta(E-\omega)}{2\omega}}\; \overline{\tilde{f}(p)}\, {\ds \frac{d\tilde{g}(p)}{\;d {\bf p}}}\\
 &\neq \langle {\bf x} f| g\rangle\\
 &= -i\hbar {\ds \int} dp \;{\ds \frac{\delta(E-\omega)}{2\omega}}\; \tilde{g}(p)\, {\ds \frac{d\overline{\tilde{f}(p)}}{\;d {\bf p}}}\end{array}\]using $\langle {\bf x} f| g\rangle =\overline{\langle g| {\bf x} f\rangle}$, $x:=t,{\bf x}$, $p:=E,{\bf p}$ and $\omega^2=m^2+{\bf p}^2$. The components of position do correspond to a self-adjoint operator in the nonrelativistic limit $m\rightarrow \infty$.\[\int dp \;{\ds \frac{\delta(E-\omega)}{2\omega}}\; \overline{\tilde{f}(p)}\, {\ds \frac{d\tilde{g}(p)}{\;d p}}\longrightarrow \int d{\bf p} \;{\ds \frac{\overline{\tilde{f}(m,{\bf p})}}{2m}}\; {\ds \frac{d\tilde{g}(m,{\bf p})}{\;d p}} =- \int d{\bf p} \;{\ds \frac{\tilde{g}(m,{\bf p})}{2m}}\; {\ds \frac{d\overline{\tilde{f}(m,{\bf p})}}{\;d p}}\]from integration by parts. For finite $m$, the self-adjoint operator that most closely corresponds with position has eigenvectors of extended spatial support [\ref{wigner},\ref{barut}]. This is another example of classical observables, in this case position, lacking an association with self-adjoint operators. While distinguishing $x$ and $p$ from products such as $x^3p$ might be acceptable in descriptions of quantum mechanics, this example demonstrates that a quantity as essential as location does not necessarily correspond directly with a self-adjoint Hilbert space operator.

The final topic of this section is that in the example of the elementary constructed QFTs with interaction [\ref{gej05}], no element in the Hilbert space is labeled by a function with support limited to a bounded subset of spacetime. This raises the possibility that there may be no Hilbert space states strictly associated with bounded regions of spacetime for fields exhibiting interaction. On this point, free fields may be misleading. Again, this lack does not exclude an observer's interpretation of states as supported in a bounded region. The Araki-Haag-Kastler algebraic QFT considers algebras of bounded, self-adjoint operators associated with bounded subsets of spacetime. The one-to-one correspondence of Hilbert space projection operators and subspaces of the Hilbert space provides that the association of Hilbert space projection operators with bounded regions of spacetime may be approximate. This approximation is a substantial distinction; consideration of the extrapolations of analytic functions that vanish in a finite subset of spacetime is very different from consideration of the extrapolations of analytic functions that approximately vanish.

% = = = = = = = = = = = = = = = = = = = = = = = = = = = = = = = = = = = = = = = =

%established description of quantum mechanics

% = = = = = = = = = = = = = = = = = = = = = = = = = = = = = = = = = = = = = = = =
\subsection*{The Everett-Wheeler-Graham description of reality}

The EWG relative state interpretation [\ref{ewg}] avoids imposition of intuition originating in classical idealizations and results from consideration that quantum mechanics is a complete and consistent model of the natural world. The interpretation is consistent with observables as descriptions of the states and there is no reliance on a conjectured classical domain. In the context of considering observable quantities like field strength when there is no association with a self-adjoint Hilbert space operator, the relative state interpretation also frees quantum mechanics from a limitation that measurement is collapse onto an eigenvector of the self-adjoint operator that corresponds to the dynamical variable.

In current understanding, the natural world is described by quantum mechanics and this description has classical limits that approximate the results of classical mechanics. This approximation is for $\hbar>0$ and whether the $\hbar \rightarrow 0$ limit of quantum mechanics should be classical mechanics is a distinct question. Although predictions are consistent, no entry into a classical domain is required. One attempt to maintain classical concepts is to consider quantum mechanics as a statistical theory for classically described states and use a discontinuous process of measurement, distinct from unitary temporal evolution [\ref{vonN}]. Then, the element of the Hilbert space describes a distribution over classical states, described as eigenvectors of the operator corresponding to the dynamical variable and the state collapses to one of those eigenvectors upon measurement. The distinctions between these two processes, collapse of the state upon measurement and unitary evolution in time, are ad hoc and inconsistent as evidenced by the Schr\"{o}dinger's cat, Wigner's friend and the Einstein-Podolsky-Rosen measurement paradoxes. A resolution to the measurement paradoxes is to accept quantum mechanics as a new description of states. The view that the states of nature truly are elements of Hilbert spaces is natural, self-consistent, and all state evolution satisfies the Schr\"{o}dinger equation. The concept that a microscopic, quantum domain interacts with a macroscopic, classical domain leads to the Schr\"{o}dinger's cat and Wigner's friend paradoxes, and considering quantum mechanics as a statistical theory for classically described objects leads to the Einstein-Podolsky-Rosen paradox. These measurement paradoxes demonstrate the inadequacies of attempts to persist in familiar, idealized classical descriptions. The incorporation of classical realms into a quantum mechanical description of nature encounters contradiction. Quantum mechanics is a shift from identification of objects by their orientations, positions and velocities to description of objects as elements in Hilbert spaces. Objects are not in classical states before or after observation. The perspective that a purely quantum mechanical description of nature is bizarre is countered by daily experience. Although it discomforts many who favor the idealized classical description of nature, the argument that quantum mechanics is a complete and consistent model for the physical world has not resulted in an observable flaw.

% = = = = = = = = = = = = = = = = = = = = = = = = = = = = = = = = = = = = = = = =

% reality of multiple worlds, virtual collapse, EPR discussion

% = = = = = = = = = = = = = = = = = = = = = = = = = = = = = = = = = = = = = = = =

Considering quantum mechanics as a complete theory leads to inclusion of descriptions for observers within state descriptions. Insight into the EWG relative state [\ref{ewg}] interpretation of quantum mechanics follows from the examination of Hilbert space structures and the concept of relative state.

A measurement is a characterization of the state observed. The record of these characterizations is part of the history of the observer. An observer's state, described by a subspace of the Hilbert space, is labeled by the record, the list of the results of all prior measurements and communications with other observers. Results of interaction are labeled by the observation, for example, the result may be perception of a point-particle at a location and this idealization may not necessarily reflect the actual description of the state. There is no assertion that the record reflects the reality.

Observer's records are never observed to be superpositions across records. This suggests that there is an unambiguous, natural basis for the observer states, one aligned with the possible records. Physical reasonableness and the separability of the Hilbert space argues that there is a partition of unity in terms of a denumerable set of orthogonal projection operators $Q_\theta$ that correspond to the subspaces of observer states labeled by the observer's record designated $\theta$. $Q_\theta^*=Q_\theta=Q_\theta^2$ and\begin{equation}\label{ptheta}\sum_\theta \; Q_\theta=1.\end{equation}No states are neglected in the summation of projections labeled by observer records. This is achievable using a decomposition of the Hilbert space into orthogonal complements [\ref{rudin}] from orthogonality of the observer states with distinct records.

Consider the conditional expectation value for any self-adjoint operator $A$. The expected value of $A$ within a subspace labeled by an observer's record $\theta$ is the conditional expectation\begin{equation}\label{rel-st}\renewcommand{\arraystretch}{2} \begin{array}{rl} E^\theta [A] &:=\frac{\ds E[Q_\theta A Q_\theta]}{\ds E[Q_\theta]}\\
  &= \frac{\ds \mbox{Trace}(Q_\theta A Q_\theta \rho)}{\ds \mbox{Trace}(Q_\theta \rho)}.\end{array}\end{equation}Following [\ref{ewg}], define $\rho^\theta$, the relative state density matrix, relative to an observer's record $\theta$, as\[\rho^\theta := c_\theta\, Q_\theta \rho Q_\theta\]with\[c_\theta^{-1} = \mbox{Trace}(Q_\theta \rho).\]The cyclic invariance of the trace and $Q_\theta^2=Q_\theta$ results in a unit trace for $\rho^\theta$. The cyclic invariance of the trace supports dual interpretations of\[E^\theta [A]=  \mbox{Trace}( A \rho^\theta).\]The relative state or EWG interpretation of quantum mechanics can be considered the result of this equivalence of conditional expectations with unconditional expectations using the relative state density matrix $\rho^\theta$. The expected value of any $A$ for the appropriate relative state density matrix equals the expected value of $A$ conditioned on knowledge of the observer's state. Characterizing this relative state is sufficient to describe the observables relative to a given history of the observer.

For any $A$ that commutes with the $Q_\theta$, that is, for any $A$ generated by projections in the commutant of the $Q_\theta$, a mixture of relative state density matrices is equivalent to the state density matrix. A weighted summation of the relative state density matrices provides the correct expectation for any observable that commutes with the projections onto the subspaces including the orthogonal states of the observer. From (\ref{ptheta}), $\sum_\theta Q_\theta =1$ is a resolution of unity in the Hilbert space. Inserting this resolution of unity, idempotence of the $Q_\theta$, noting that the $A$ and the $Q_\theta$ commute, the cyclic invariance of the trace and the definition of relative state provides the equivalence.\[\renewcommand{\arraystretch}{1.25} \begin{array}{rl} E[A]&=\mbox{Trace}( A \rho)\\
&= {\ds \sum_\theta}\;  \mbox{Trace}( Q_\theta A \rho)\\
&= {\ds \sum_\theta} \; \mbox{Trace}( Q_\theta^2 A \rho)\\
&= {\ds \sum_\theta} \; \mbox{Trace}( Q_\theta A Q_\theta \rho)\\
&= {\ds \sum_\theta}\; \mbox{Trace}( A Q_\theta \rho Q_\theta)\\
&= {\ds \sum_\theta}\; \mbox{Trace}( A \rho^\theta)/c_\theta. \end{array}\]Then the mixture of relative states,\[\rho^{\mathit{eq}}:= {\ds \sum_\theta}\; \rho^{\theta}/c_\theta\]describes the same observables relative to the observer as the complete description $\rho$.\[\mbox{Trace}(A \rho) =\mbox{Trace}(A \rho^{\mathit{eq}})\]for all $A$ in the commutant of the $Q_\theta$.

With the observer as well as the system under observation in the quantum mechanical description, time evolution is continuous, unitary time translation P2. The equivalence of $\rho^{\mathit{eq}}$ and $\rho$ provides that there is no need to define what interactions constitute measurements nor a need for a collapse of the system state upon measurement. Measurement can be distinguished as any interaction that results in change to the observer's state. The observable equivalence of states is the reason that only the history relative to one observer record needs to be maintained. An observer need not carry the entire history of possibilities, the state density matrix $\rho$, to describe the natural world. All future observations are relative to the current state of the observer, that is, they are conditioned upon a particular record $\theta$, and a description of that relative state is all that must be maintained to properly describe the future. With attention focused on one term in the mixture of relative states, this term can be interpreted as the result of an effective, virtual collapse to that state upon measurement. The state density matrix continues in a temporally continuous evolution. Since the mixture of relative states is physically equivalent to the complete description of the natural world, limiting knowledge to the relative state does not modify the observables of the system. The act of measurement defines the state of the observer as well as the relative state of the system. The system is left in a state correlated with the observer's knowledge. Good measurements convey accurate descriptions of the relative state of the system under observation. There is no consequence to considering the alternative states since their inclusion does not effect further observation.

Whether the other ``branches'' are real is a metaphysical issue of no empirical consequence. There is no observable effect whether the branches are considered or not. The evolution of the state is indifferent to whether the branching or collapsing is considered. There is an equivalences of: one complete description that includes the observer distributed over possibilities; a description of many, autonomous worlds labeled by the possible results recorded by the observer; and a description that collapses to the state relative to one classically described observer upon each measurement. Results are indifferent to the approach considered. All three are within quantum mechanics considered as a complete description. The concepts of classical physics are poorly posed for the description of quantum mechanics.

The observation about quantum mechanics captured in the Einstein-Podolsky-Rosen (EPR) paradox might be described as that quantum mechanics violates the intuition originating in classical mechanics, as it surely does. A classical description of states is inconsistent with nature. A close association of classical concepts with quantum mechanics, canonical quantization, suffices to a limited extent in ordinary quantum mechanics but additional difficulty arises with the union of relativity and quantum mechanics.

\subsection*{Measurement and EPR}

In this section, an explicit, simplified example of measurement is discussed with the perspective that quantum mechanics is a complete description of nature. In this development, there is no need to distinguish measurement as an interaction other than that measurements produce changes to an observer's state, and no need for wave packet collapse or classical domains. The EPR paradox is used to emphasize essential differences between classical and quantum mechanical state descriptions. 

The interaction of an observer with a subsystem, for this illustration taken to be the two spin polarization states of a spin one-half elementary particle, is a measurement. This measurement may be perfect, any projection onto the observer state that records a spin up state includes only spin up subsystem states, or not, a projection onto the observer state that records a spin up state includes a mixture of spin up and spin down subsystem states. Three orthogonal states of the observer are distinguished in this example: the observer has no record for the spin; the observer recorded spin up; or the observer recorded spin down. Whether the elementary system is actually spin up when the observer records spin up relies on experiment design and in the case of more general measurements, the structure of the Hilbert space. Indeed, in the case of position, there are no orthogonal projection operators associated with a point in a Poincar\'{e} covariant one particle subspace.

The Hilbert space is designated ${\bf H}$ and, in this simple example, is decomposed into three orthogonal subspaces labeled by the result of observation. The appeal is to physical reasonableness that the three labeled states of the observer can be taken to be orthogonal. If the observing system is sufficiently complex to have three linearly independent states then an orthogonal set can be constructed by Gram-Schmidt construction. An appeal to determinism fixes the three states labeled by the three possible results of measurement as orthogonal. It is not reasonable to anticipate any likelihood that an observer will change the record subsequent to observation, for example, upon inquiry by Wigner's friend.

Projections onto the three orthogonal subspaces are labeled $Q_\mathit{xx}$, $Q_\mathit{up}$ and $Q_\mathit{dn}$. Without loss of generality it can be assumed that the projections are a decomposition of unity,\[Q_\mathit{up}+Q_\mathit{dn}+Q_\mathit{xx}=1.\]This correponds to decomposing the Hilbert space into three orthogonal complements [\ref{rudin}] and ensuring that each of the subspaces include only one of the three orthogonal states of the observer. Then the subspaces are labeled by the observer's record: $\mathit{xx}$ is no observation, $\mathit{up}$ is spin up recorded and $\mathit{dn}$ is spin down recorded. The three orthogonal subspaces are designated ${\bf H}_\mathit{up}$, ${\bf H}_\mathit{dn}$ and ${\bf H}_\mathit{xx}$ with ${\bf H}= {\bf H}_\mathit{up}\oplus {\bf H}_\mathit{dn} \oplus {\bf H}_\mathit{xx}$. There is no reason to believe that the structures of the three subspaces differ except by the observer's record and relative state. Consequently assert that two self-adjoint projections $P_k=P_k^*=P_k^2$ for $k=1,2$ and a self-adjoint Hamiltonian operator $H_o$ are defined in all three subspaces. The projection operators are for the spin polarizations of the subsystem, $P_1$ for spin up and $P_2$ for spin down. Assuming a rotationally invariant Hamiltonian $H_o$, the Hamiltonian simultaneously diagonalizes with the projection operators, $[H_o,P_k]=0$ for $k=1,2$. Also require that the $P_k$ are projections onto orthogonal complements in the ${\bf H}_a$, $P_1+P_2=1$.

The interaction that results in the measurement is defined by an interaction Hamiltonian that applies for a finite interval, $0\leq t \leq T_m$. This is one of several idealizations. More typically, the interaction will apply for any finite separation of states, an interaction that only approximately vanishes for finite times with states that lack bounded support. To define the Hamiltonian, set\[\renewcommand{\arraystretch}{1.25} \begin{array}{c} A_1 := \cos\vartheta P_1 +\sin\vartheta P_2\\
A_2 := \cos\vartheta P_2 +\sin\vartheta P_1\end{array}\]using the orthogonal projection operators. The self-adjointness, orthogonality and idempotence of the projection operators provide that $A_k^*=A_k$, $[A_1,A_2]=0$ and\[A_1^2+A_2^2=1.\]The Hamiltonian on the interval $0 \leq t \leq T_m$ is set to\[H=\frac{\pi}{2 T_m} \left( \begin{array}{ccc}0& 0& i A_1\\ 0& 0& i A_2\\ -i A_1& -i A_2& 0\end{array} \right) +\left( \begin{array}{ccc} H_0& 0& 0\\ 0& H_0& 0\\ 0& 0& H_0\end{array} \right)\]with\[H=\left( \begin{array}{ccc} H_0& 0& 0\\ 0& H_0& 0\\ 0& 0& H_0\end{array} \right)\]otherwise. The Hilbert space is composed as the direct sum of the orthogonal subspaces in the order ${\bf H}_\mathit{up}\oplus {\bf H}_\mathit{dn} \oplus {\bf H}_\mathit{xx}$. Three sets of orthonormal eigenvectors for $H$ are\[e_0=\left( \begin{array}{r} A_2 w_0\\ -A_1 w_0\\ 0\quad \end{array} \right),\qquad e_\pm=\frac{1}{\sqrt{2}}\left( \begin{array}{c} A_1 w_\pm\\ A_2 w_\pm\\ \pm i w_\pm\end{array} \right)\]with $w$ that are eigenvectors of $H_0$ with eigenvalues $E_w$ and $\|w\|=1$. The corresponding eigenvalues are $\lambda_0=E_{w_0}$ and $\lambda_\pm =\pm\, \pi/(2T_m)+E_{w_\pm}$.

With $R_a$ designating the projections onto subspaces spanned by the three sets of eigenvalues labeled $a=0,\pm$, the time translation operator in ${\bf H}$ is\[\renewcommand{\arraystretch}{1.25} \begin{array}{rl}U(t)&= U_0(t)\,(e^{i\phi} R_+ +e^{-i\phi} R_- +R_0)\\&= U_0(t) \left( \renewcommand{\arraystretch}{1} \begin{array}{ccc} A_2^2 +A_1^2 \cos \phi&  A_1A_2\, (\cos \phi-1) & A_1 \sin \phi\\  A_1A_2\, (\cos \phi-1)& A_1^2 +A_2^2 \cos \phi& A_2 \sin \phi\\ -A_1 \sin \phi& -A_2 \sin \phi&  \cos \phi\end{array} \right)\end{array}\]with a common factor on matrix elements of\[U_0(t):=e^{iH_0 t}\]and\[\phi:= \left\{ \renewcommand{\arraystretch}{1} \begin{array}{ll}0 &t<0\\
\pi/2 & t>T_m\\
\pi t/(2T_m)\qquad &\mbox{otherwise.}\end{array}\right.\]An initial state\[\rho(0):=\left( \begin{array}{ccc}0& 0& 0\\ 0& 0& 0\\ 0& 0& \rho_0\end{array} \right)\]evolves to\[\renewcommand{\arraystretch}{1.25} \begin{array}{rl}\rho(t)&=U(t) \rho(0)U(t)^{-1}\\
 &=\left( \begin{array}{ccc} A_1 \rho_0(t) A_1 \sin^2\phi&  A_1 \rho_0(t) A_2 \sin^2\phi&  A_1 \rho_0(t) \sin\phi \cos\phi\\
A_2 \rho_0(t) A_1 \sin^2\phi&  A_2 \rho_0(t) A_2 \sin^2\phi&  A_2 \rho_0(t) \sin\phi \cos\phi\\
\rho_0(t) A_1 \sin\phi \cos\phi &  \rho_0(t) A_2 \sin\phi \cos\phi& \rho_0(t) \cos^2\phi\end{array} \right)\end{array}\]for $t>0$ with $\rho_0(t):=U_0(t)\rho_0U_0(t)^{-1}$ and in particular,\[\rho(t)=\left( \begin{array}{ccc} A_1 \rho_0(t) A_1&  A_1 \rho_0(t) A_2&  0\\
A_2 \rho_0(t) A_1&  A_2 \rho_0(t) A_2& 0\\
0 & 0& 0\end{array} \right)\]when $t>T_m$.

With\[\hat{P}_k:=\left( \begin{array}{ccc} P_k& 0& 0\\ 0& P_k& 0\\ 0& 0& P_k\end{array}\right),\]the likelihood that the system of interest is spin up for $t\leq 0$ is Trace$(\hat{P}_1 \rho)=$Trace$(P_1 \rho_0)$ and the conditional likelihood for spin up, conditioned on the observer's record, is\[L_{1|a}=\frac{\mbox{Trace}(\hat{P}_1 Q_a \rho\, Q_a)}{\mbox{Trace}(Q_a \rho)}.\]From the selection Trace$(P_1 \rho_0)=\frac{1}{2}$, polarization of the spin is equally likely to be up or down initially, and the likelihood of the observer's record $a$ is\[\mbox{Trace}(Q_a \rho)= \left\{ \begin{array}{rl} \frac{1}{2} \sin^2\phi &\qquad \mbox{for } a=\mathit{up}\mbox{ or }\mathit{dn}\\
 \cos^2\phi & \qquad \mbox{for } a=\mathit{xx}.\end{array}\right.\]

The quality of the measurement is assessed by examining the conditional expectations of spin up\[\frac{\mbox{Trace}(\hat{P}_1 Q_a \rho\, Q_a)}{\mbox{Trace}(Q_a \rho)}= \left\{ \begin{array}{ll} \cos^2\vartheta \qquad& \mbox{for } a=\mathit{up}\\
 \sin^2\vartheta & \mbox{for } a=\mathit{dn}\\
 \frac{1}{2} & \mbox{for } a=\mathit{xx}\end{array}\right.\]and spin down\[\frac{\mbox{Trace}(\hat{P}_2 Q_a \rho\, Q_a)}{\mbox{Trace}(Q_a \rho)}= \left\{ \begin{array}{ll} \sin^2\vartheta \qquad& \mbox{for } a=\mathit{up}\\
 \cos^2\vartheta & \mbox{for } a=\mathit{dn}\\
 \frac{1}{2} & \mbox{for } a=\mathit{xx}.\end{array}\right.\]The example illustrates that any interaction that changes the observer's state is a measurement. The measurement may be perfect ($\vartheta=k\pi$), may result in no correlation of observer with observed states ($\vartheta=(4k+1)\pi/4, (4k+3)\pi/4$), result in the wrong correlation of spin with record ($\vartheta=(2k+1)\pi/2$), or any variation between these extremes. The measurement may also leave a finite likelihood that the observer records no observation, in the example when the interaction Hamiltonian applies only for $t<T_m$ ($\phi < \pi/2$).

The EPR paradox, whether the quantum description of a state has reality if the actions of a distant observer can apparently affect the description of the state, demonstrates that a classically idealized description is inconsistent with nature. When two spin one-half particles fly apart, a classical state of determined spin polarization can not be attributed to each of the particles. The spin polarizations are correlated and angular momentum conserved, no matter how great the separation of the product particles. The actions of an observer interacting with one of the two product particles interacts with the joint state that includes description of the distant state. In the example above, initially the observer is uncorrelated with the state of the two particles. The interaction produces correlation of the observer's state with the polarization of one particle, spin up or down, and that spin state remains correlated with the distant paired spin state. If the observer chooses to measure spin on a different axis, the interaction would be different but the spin polarizations remain in correlated pairs that conserve angular momentum. The paradox derives from attributing a classically idealized state to the distant particle.

While the example suffers from several idealizations, a measurement process is illustrated. The idealizations include the finite duration of the interaction, and that both the observer and observed subsystem have a finite number of linearly independent states. The remainder of the ${\bf H}_a$ may have more general structure.

% = = = = = = = = = = = = = = = = = = = = = = = = = = = = = = = = = = = = = = = =

% conclusion

% = = = = = = = = = = = = = = = = = = = = = = = = = = = = = = = = = = = = = = = =
\section{Discussion}

Of the three revolutions in physics of the early 20th century, special and general relativity remained classical theories, that is, descriptions for objects moving along trajectories in spacetime and for fields that are functions on spacetime. Quantum mechanics departed from the classical concepts for particles and fields, and described physical states as elements of Hilbert spaces. These Hilbert space descriptions lack many properties of the classical idealizations, in particular, the descriptions can lack evident interpretations as geometric objects. Although much of the development of quantum mechanics has been derivations for quantum mechanics from classical mechanics and classical field theory, even the concept of point particle is unnatural to relativistic quantum mechanics. Canonical quantizations are conjectures that translate classical dynamical variables to self-adjoint Hilbert space operators to describe ``quantum'' effects. This effort is despite evident contradictions, such as the EPR paradox, to the classical idealizations. While Ehrenfest's theorem provides justification for a correspondence of dynamical variables in nonrelativistic approximations when particle production is precluded, a correspondence at high energies is a more speculative extrapolation. Assertion that observables of interest are necessarily self-adjoint Hilbert space operators contradicts the knowledge that such a correspondence is not the general case and that the eigenstates of self-adjoint operators can be associated with generalized functions as well as states. The difficulties of Lagrangian QFT may derive from this unnaturally limited description for relativistic quantum dynamics. The difficulties include description of quantum mechanics at relativistic energies using operators conjectured to be realized in Hilbert spaces of interest but with inconsistent properties. As the dynamical variable $x^3p$ for the linear harmonic oscillator demonstrates, approximation of classical mechanics by quantum mechanics is richer than a correspondence of classical dynamical variables with self-adjoint operators. Classical limits of the harmonic oscillator accurately approximate the classical dynamics of $x^3p$ even though $x^3p$ does not correspond to a self-adjoint Hilbert space operator. And, no appeal to a classical domain nor collapse to an eigenfunction of $x^3p$ is required to observe $x^3p$. Hermiticity of densely defined Hilbert space field operators has been considered a foundation principle of QFT even though examples illustrate that classical dynamical variables need not correspond to Hermitian operators for quantum dynamics to provide an accurate approximation of classical dynamics for appropriate states. Classical fields may be classical limits of quantum mechanical descriptions that lack self-adjoint field operators.

% = = = = = =
% = = = = = =
% = = = = = =
\section{References}
\begin{enumerate}
\item \label{longo} R.~Brunetti, D.~Guido, and R.~Longo, ``Modular Localization and Wigner Particles'',  {\em Rev.~Math.~Phys.}, Vol.~14, 2002, p. 759.
\item \label{ewg} B.S.~DeWitt, H.~Everett III, N.~Graham, J.A.~Wheeler, et al republished in {\em The Many-worlds Interpretation of Quantum Mechanics}, Ed. B.S.~DeWitt, N. Graham, Princeton Univ.~Press, Princeton, N.J., 1973.
\item \label{rudin} Walter Rudin, {\em Real \& Complex Analysis}, New York, NY: McGraw-Hill Book Co., 1966.
\item \label{gel4} I.M.~Gel'fand, and N.Ya.~Vilenkin, {\em Generalized Functions}, Vol.~4, trans.~A.~Feinstein, New York, NY: Academic Press, 1964.
\item \label{yngvason-lqp} J.~Yngvason, ``Localization and Entanglement in Relativistic Quantum Physics'', 12 January, 2014, arXiv:quant-ph/1401.2652v1.
\item \label{gej05} G.E.~Johnson, ``Algebras without Involution and Quantum Field Theories'', 13 March, 2012, arXiv:math-ph/1203.2705v2.
\item \label{mp01} G.E.~Johnson, ``Massless Particles in QFT from Algebras without Involution'', 22 May, 2012, arXiv:math-ph/1205.4323v2.
\item \label{feymns} G.E.~Johnson, ``Fields and Quantum Mechanics'', 17 Dec.~2013, arXiv:math-ph/1312.\-2608v4.
\item \label{dirac} P.A.M.~Dirac, {\em The Principles of Quantum Mechanics, Fourth Edition}, Oxford: Clarendon Press, 1958.
\item \label{bjdrell} J.D.~Bjorken and S.D.~Drell, {\em Relativistic Quantum Mechanics}, New York, NY: McGraw-Hill, 1964.
%\item \label{bjdrell2} J.D.~Bjorken and S.D.~Drell, {\em Relativistic Quantum Fields}, New York, NY: McGraw-Hill, 1965.
\item \label{weinberg} S.~Weinberg, {\em The Quantum Theory of Fields, Volume I, Foundations}, New York, NY: Cambridge University Press, 1995.
\item \label{vonN} J.~von Neumann, {\em Mathematical Foundations of Quantum Mechanics}, Princeton, NJ: Princeton University Press, 1955.
\item \label{gleason} A.M.~Gleason, ``Measures on the closed subspaces of a Hilbert space'', {\em Journal of Mathematics and Mechanics}, Vol.~6, 1957, p.~ 885–893.
\item \label{wigner} T.D.~Newton and E.P.~Wigner, ``Localized States for Elementary Systems'', {\em Rev. Modern Phys.}, Vol.~21, 1949, p.~400.
\item \label{wizimirski} Z.~Wizimirski, ``On the Existence of a Field of Operators in the Axiomatic Quantum Field Theory'', {\em Bull. Acad. Polon. Sci., s\'{e}r. Math., Astr. et Phys.}, Vol.~14, 1966, pg.~91.
\item \label{wight-pt} A.S.~Wightman, ``La th\'{e}orie quantique locale et la  th\'{e}orie quantique des champs'', {\em Ann. Inst. Henri Poincar\'{e}}, Vol.~A1, 1964, p.~403.
\item \label{segal} I.E.~Segal and R.W,~Goodman, ``Anti-locality of certain Lorentz-invariant operators'', {\em Journal of Mathematics and Mechanics}, Vol.~14, 1965, p.~629.
\item \label{gel2} I.M.~Gel'fand, and G.E.~Shilov, {\em Generalized Functions}, Vol.~2, trans.~M.D.~Friedman, A.~Feinstein, and C.P.~Peltzer, New York, NY: Academic Press, 1968.
\item \label{birk} G.~Birkhoff and J.~von Neumann, ``The Logic of Quantum Mechanics'', {\em Ann. Math.}, Vol.~37, 1936, pp. 823-842.
\item \label{bogo} N.N.~Bogolubov, A.A.~Logunov, and I.T.~Todorov, {\em Introduction to Axiomatic Quantum Field Theory}, trans.~by Stephen Fulling and Ludmilla Popova, Reading, MA: W.A.~Benjamin, 1975.
\item \label{schrodinger} E. Schr\"{o}dinger, ``Der stetige Übergang von der Mikro-zur Makromechanik'', {\em Die Naturwissenschaften}, Vol.~14. Issue~28, 1926, p.~664.
\item \label{barut} A.O.~Barut and S.~Malin, ``Position Operators and Localizability of Quantum Systems Described by Finite- and Infinite-Dimensional Wave Equations'', {\em Rev. Modern Phys.}, Vol.~40, 1968, p.~632.
\end{enumerate}
\end{document}